\documentclass[12pt,thmsa]{article}

\usepackage{cite}
\usepackage{graphicx}

\begin{document}

\title{Rayleigh--Ritz variation method and connected--moments polynomial approach}
\author{Francisco M. Fern\'{a}ndez \thanks{%
e--mail: fernande@quimica.unlp.edu.ar} \\
INIFTA (UNLP, CCT La Plata--CONICET), \\
Divisi\'{o}n Qu\'{i}mica Te\'{o}rica,\\
Diag. 113 y 64 (S/N), Sucursal 4, Casilla de Correo 16,\\
1900 La Plata, Argentina}
\maketitle

\begin{abstract}
We show that the connected--moments polynomial approach proposed recently is
equivalent to the well known Rayleigh--Ritz variation method in the Krylov
space. We compare the latter with one of the original connected--moments
methods by means of a numerical test on an anharmonic oscillator.
\end{abstract}

\section{Introduction}

The t--expansion\cite{HW84} has motivated a great interest in the
application of the connected--moments expansion (CMX)\cite{C87} and its
variants\cite{K87,S88,MZM94} to quantum--mechanical models. In spite of its
limitations\cite{K87,S88,MZM94,MBM89,MPM91,MZMMP94} the CMX has proved
useful for the study of many physical problems\cite
{C87,K87,MBM89,MPM91,C87b,C87c,C87d,FMB02}.

Recently, Bartashevich\cite{B08} proposed the connected--moments polynomial
approach (CMPA) that yields approximate eigenvalues to all states as roots
of a simple polynomial function of the energy with coefficients that depend
on the moments of the Hamiltonian operator.

The purpose of this paper is to show that the CMPA is simply the
Rayleigh--Ritz variation method (RRVM)\cite{MD33} in a Krylov subspace.

\section{Rayleigh--Ritz variation method and connected--moments polynomial
approach}

It is our purpose to solve the eigenvalue equation
\begin{equation}
\hat{H}\left| \psi _{j}\right\rangle =E_{j}\left| \psi _{j}\right\rangle
,\;j=0,1,\ldots  \label{eq:eigen}
\end{equation}
by means of the RRVM\cite{MD33} in the Krylov space spanned by $\left\{
\left| \phi _{j}\right\rangle =\hat{H}^{j}\left| \phi \right\rangle \right\}
_{j=0}^{\infty }$, where $\left| \phi \right\rangle $ is a properly chosen
vector of the state space. The RRVM makes the trial function
\begin{equation}
\left| \varphi \right\rangle =\sum_{j=0}^{N-1}c_{j}\left| \phi
_{j}\right\rangle  \label{eq:trial}
\end{equation}
orthogonal to the subspace spanned by $\left\{ \left| \phi _{j}\right\rangle
=\hat{H}^{j}\left| \phi \right\rangle \right\} _{j=0}^{N-1}$; that is to
say:
\begin{equation}
\left\langle \phi _{j}\right| (\hat{H}-W)\left| \varphi \right\rangle
=0,\;j=0,1,\ldots ,N-1  \label{eq:orthogonality}
\end{equation}
where $W$ is one of the RRVM roots and an approximation to the corresponding
eigenvalue. The coefficients $c_{j}$ are solutions to the secular equations
\begin{equation}
\sum_{i=0}^{N-1}\left( H_{ji}-WS_{ji}\right) c_{i}=0,\;j=0,1,\ldots ,N-1
\label{eq:secular}
\end{equation}
where the matrix elements $H_{ji}=\left\langle \phi _{j}\right| \hat{H}%
\left| \phi _{i}\right\rangle =\mu _{i+j+1}$ and $S_{ji}=\left\langle \phi
_{j}\right| \left. \phi _{i}\right\rangle =\mu _{i+j}$ are given in terms of
the moments $\mu _{j}=\left\langle \phi \right| \hat{H}^{j}\left| \phi
\right\rangle $. There are nontrivial solutions to the homogeneous system of
linear equations (\ref{eq:secular}) only for the $N$ values of $%
W=W_{0},W_{1},\ldots ,W_{N-1}$ that are roots of the secular determinant
\begin{equation}
\left| \mu _{i+j+1}-W\mu _{i+j}\right| _{i,j=0}^{N-1}=0.
\label{eq:secular_det}
\end{equation}

It is well known that the RRVM approximate eigenvalues approach the exact
ones from above: $W_{n}^{[N]}>W_{n}^{[N+1]}>E_{n}$, $n=0,1,\ldots ,N-1$\cite
{MD33}. This statement is not valid if by chance $\left\langle \phi \right|
\left. \psi _{i}\right\rangle =0$ because $\left\langle \phi _{j}\right|
\left. \psi _{i}\right\rangle =0$, $j=0,1,\ldots $, and the eigenvector $%
\left| \psi _{i}\right\rangle $ is bypassed by the method because it is
orthogonal to the Krylov space. Here we assume that the RRVM converges,
criteria for the convergence of the approach have been discussed elsewhere%
\cite{KB77}.

It follows from (\ref{eq:orthogonality}) that $\left\langle \phi _{j}\right|
\left. \varphi \right\rangle =W^{j}\left\langle \phi \right| \left. \varphi
\right\rangle $, which, by virtue of (\ref{eq:trial}), enables us to write
\begin{equation}
\left\langle \phi \right| \left. \varphi \right\rangle
\sum_{j=0}^{N}p_{j}W^{j}=\sum_{i=0}^{N-1}c_{i}\left( \sum_{j=0}^{N}p_{j}\mu
_{i+j}\right)  \label{eq:RRCMX1}
\end{equation}
and we conclude that if the coefficients $p_{j}$ satisfy
\begin{equation}
\sum_{j=0}^{N}p_{j}\mu _{i+j}=0,\;i=0,1,\ldots ,N-1  \label{eq:pj_mui}
\end{equation}
then
\begin{equation}
\sum_{j=0}^{N}p_{j}W^{j}=0.  \label{eq:characteristic}
\end{equation}
It is clear that we can arbitrarily choose $p_{0}=1$ and that the remaining
coefficients $p_{j}$, $j=1,2,\ldots ,N$ are identical to the $x_{N-j+1}$ of
Bartashevich\cite{B08}.

Since equation (\ref{eq:characteristic}) is satisfied by all the RRVM roots,
we conclude that it is exactly the characteristic polynomial that results
from the secular determinant (\ref{eq:secular_det}). In other words, the
CMPA is equivalent to the RRVM which explains why the CMPA sequences
converge from above\cite{B08}.

In our opinion, solving the well known determinantal equation (\ref
{eq:secular_det}) is probably easier than solving equations (\ref{eq:pj_mui}%
) and (\ref{eq:characteristic}) of Bartashevich's CMPA.

If we consider the $N$ equations (\ref{eq:pj_mui}) plus (\ref
{eq:characteristic}) as a whole system of $N+1$ equations with $N+1$
unknowns $p_{0},p_{1},\ldots ,p_{N}$, then we conclude that there will be
nontrivial solutions provided that
\begin{equation}
\left|
\begin{array}{cccc}
\mu _{0} & \mu _{1} & \cdots & \mu _{N} \\
\mu _{1} & \mu _{2} & \cdots & \mu _{N+1} \\
\vdots & \vdots & \ddots & \vdots \\
\mu _{N-1} & \mu _{N} & \cdots & \mu _{2N-1} \\
1 & W & \cdots & W^{N}
\end{array}
\right| =0  \label{eq:Rafa}
\end{equation}
which is exactly the determinantal equation derived earlier by Bishop et al%
\cite{BFBG89}.

\begin{figure}[H]
\begin{center}
\includegraphics[width=9cm]{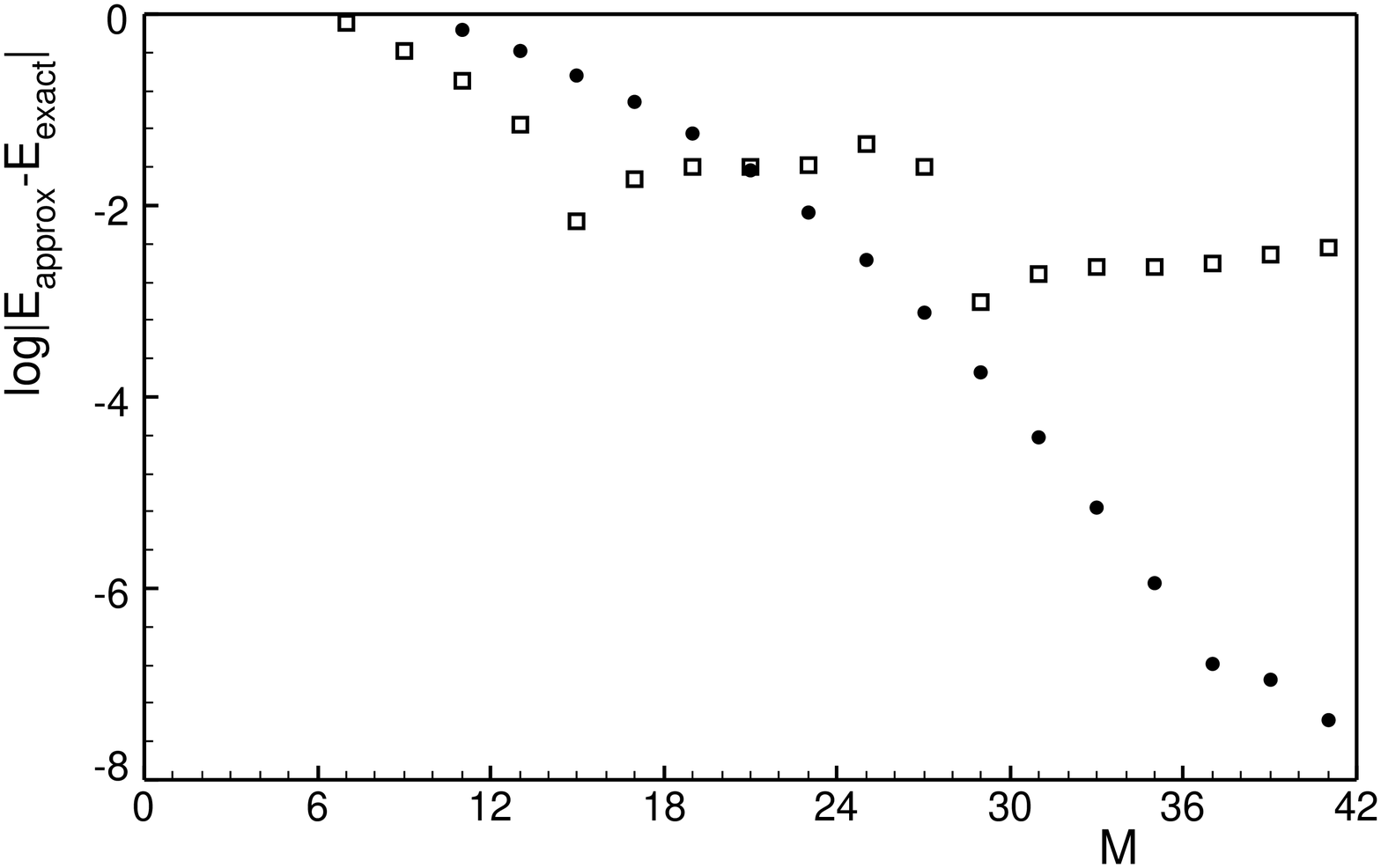}
\end{center}
\caption{Rate of convergence measured as $log|E_{approx}-E_{exact}|$ for the
CMX-LT (squares) and RRVM (filled circles) in terms of the required number
of moments $M$}
\label{fig:convergence}
\end{figure}

It is interesting to compare the original CMX approaches with the
Rayleigh--Ritz variation method in the Krylov space discussed above. For
concreteness, we arbitrarily choose the so--called CMX-LT\cite{K87}. Fig~\ref
{fig:convergence} shows $\log \left| E_{approx}-E_{exact}\right| $ for the
ground state of the dimensionless strongly--anharmonic oscillator $\hat{H}%
=-d^{2}/dx^{2}+x^{8}$. In both cases we chose the trial function $%
\left\langle x\right. \left| \phi \right\rangle =e^{-ax^{2}}$, where $a=8$,
and calculated the ``exact'' eigenvalue $E_{0}=1.225820113800492191$ by
means of the Riccati--Pad\'{e} method that exhibits remarkable rate of
convergence for this kind of models\cite{FMT89b}. We appreciate that the
RRVM converges smoothly from above while the CMX--LT exhibits great
oscillation. The same situation takes places for other values of $a$.

\section{Conclusions}

We have shown that the CMPA is equivalent to the RRVM. Consequently it
applies to all the states and produces sequences of roots that converge from
above. Numerical difficulties in applications of the CMX have been
attributed to singularities in the Hankel matrix of connected moments which
may not be invertible\cite{S88}. On the other hand, the Hankel matrix of
moments that appear in the calculation of the coefficients $x_{j}$\cite{B08}
(or present $p_{j}$'s given by (\ref{eq:pj_mui})) is always invertible. This
is not surprising because one does not expect the RRVM to exhibit such kind
of singularities. A numerical test on a simple anharmonic oscillator
suggests that the RRVM may be more reliable that the CMX approaches.

\end{document}